\documentclass[12pt]{spieman}  % 12pt font required by SPIE;

\usepackage{amsmath,amsfonts,amssymb}
\usepackage{graphicx}
\usepackage{setspace}
\usepackage{tocloft}
\usepackage{lineno}
\title{Studying the Dust Distribution Around Accreting Black Holes with
Reverberation Mapping Using PRIMA}

\author[a]{Varoujan Gorjian}
\author[a]{Michael W. Werner}
\affil[a]{Jet Propulsion Laboratory, California Institute of Technology, 4800 Oak Grove Drive, Pasadena, CA 91109, USA}
\cftpagenumbersoff{figure}
\cftpagenumbersoff{table} 
\begin{document} 
\maketitle

\begin{abstract}

Variability studies are a powerful tool for studying the structures of unresolved sources. One such type of variability study, called reverberation mapping (RM), established that the dominant source of infrared radiation from an active galactic nucleus (AGN) was from dust absorption and re-emission, which demonstrated that the optical brightening and fading of light from the accretion disk (AD) around a supermassive black hole was followed by a corresponding (delayed) variation at infrared (IR) wavelengths from the surrounding dust distribution. Since that time a great deal more has been learned about the dust distribution around ADs, both near the AD in the form of a potential torus, as well as extended emission in the form of polar dust outflows. Understanding the dust distribution is vital to understanding how AGNs affect their host galaxy as well as the overall energetics of AGNs as $\sim$50\% of energy from an AGN comes out in the IR. Dust RM has been done exclusively in the near-IR (1-5 $\mu$m) which traces the inner edge of the dust near the dust sublimation radius. Hence extending RM to the mid-IR, especially to the peak of the dust emission between 25 and 30 $\mu$m, allows for an examination of the dust distribution around ADs and potentially traces the source of the polar outflow. RM with the proposed PRobe far-Infrared Mission for Astrophysics (PRIMA) can focus on variability-monitoring of a sample of low- to high-luminosity AGNs to trace the 25-30 $\mu$m emission that is reverberated from the UV/optical AD emission which will be monitored by other space and ground-based observatories and made available to PRIMA users. Then, by detailed modeling of the response of the dust emission to shorter wavelengths, the distribution of the dust around an AGN will be revealed and can be linked to the accretion disk luminosities.

\end{abstract}

% Include a list of up to six keywords after the abstract

% Include email contact information for corresponding author
{\noindent \footnotesize\textbf{*}Varoujan Gorjian,  \linkable{vg@jpl.nasa.gov} }
\smallskip

© 2025 California Institute of Technology. Government sponsorship acknowledged

\begin{spacing}{2}   % use double spacing for rest of manuscript

\section{Introduction}
\label{sect:intro}  % \label{} allows reference to this section
Variability studies are a powerful tool for studying the structures of unresolved sources. One such type of variability study, called reverberation mapping (RM), established that by monitoring the variability of active galactic nuclei (AGN) where gas is accreting onto a supermassive black hole (SMBH) at the center of a galaxy, could lead to an understanding of the size and extent of the accretion disk and its surroundings. \cite{Blandford..1982,Peterson..1993,Kriss.etal.2000,Sergeev.etal.2005,Shappee.etal.2014}.

\begin{figure}
\begin{center}
%\begin{tabular}{c}
\includegraphics[height=9cm]{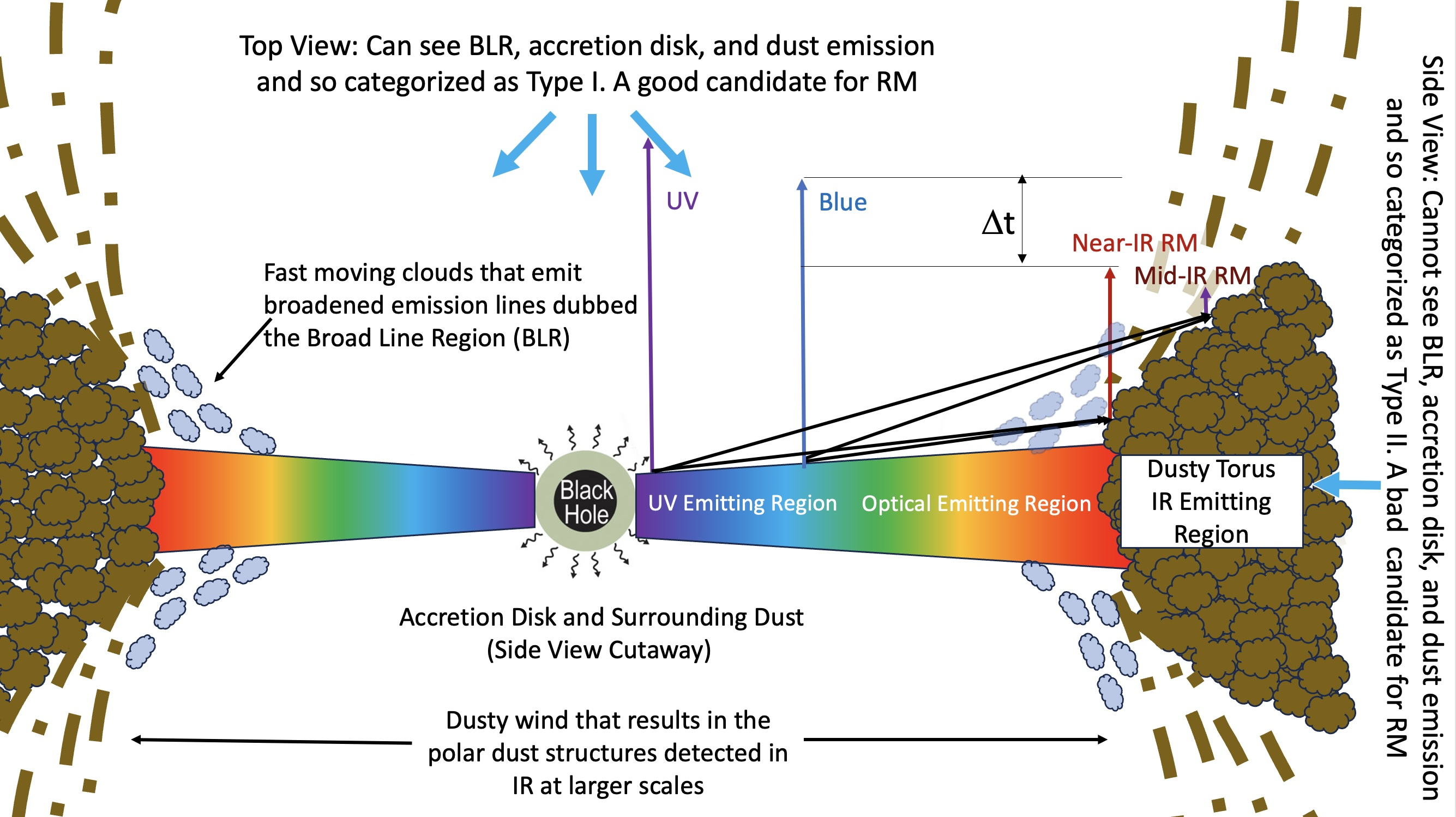}  
%\hspace{6in}
%\end{tabular}
\end{center}
\caption 
{ \label{fig:AGN-Schematic}
A schematic (not to scale) representation of reverberation mapping. The differing arrival times of differing wavelengths (UV/optical vs IR) to an observer from the top correspond to the physical separation at a length of c$\Delta$t between those regions where $\Delta$t is the difference in arrival times of differing wavelengths. To guarantee that the dust does not block the UV/optical emitting region, only Type I AGN are selected for RM where the presence of broad lines in the AGN spectrum guarantee an unobstructed view of the BLR and hence the accretion disk. Adapted from  Ref[\citenum{CackettBentzKara2021}]} 
\end{figure}

Earlier RM efforts concentrated on the spectroscopic emission from fast-moving gas in the broad line region (BLR) of AGN, Figure \ref{fig:AGN-Schematic}, but continuum RM was subsequently used to show that a significant source of near-infrared radiation from AGN came from dust absorption and re-emission of short wavelength light generated closer to the black hole, where UV/optical brightening and fading was followed by a corresponding (delayed) variation at infrared (IR) wavelengths \cite{Clavel..etal..1989,Nelson..1996} Also, in the simplest version of how the dust is distributed, a thick dust torus determines whether a viewer sees a broad-lined/unobscured Type I AGN or a narrow-lined/obscured Type II AGN\cite{Antonucci93}. Hence by choosing Type I AGN which show broad emission lines in their spectra, RM observations guarantee that the dusty torus is not impeding the observer's view of the region near the SMBH and the accretion disk and that the observations are tracing a radial distribution outward form the SMBH. Type II AGN, which do not show broad emission lines in their spectra, are presumed to be viewed edge-on with the dust blocking the BLR, and so are not chosen for RM campaigns.

\begin{figure}
\begin{center}
%\begin{tabular}{c}
\includegraphics[height=7.5cm]{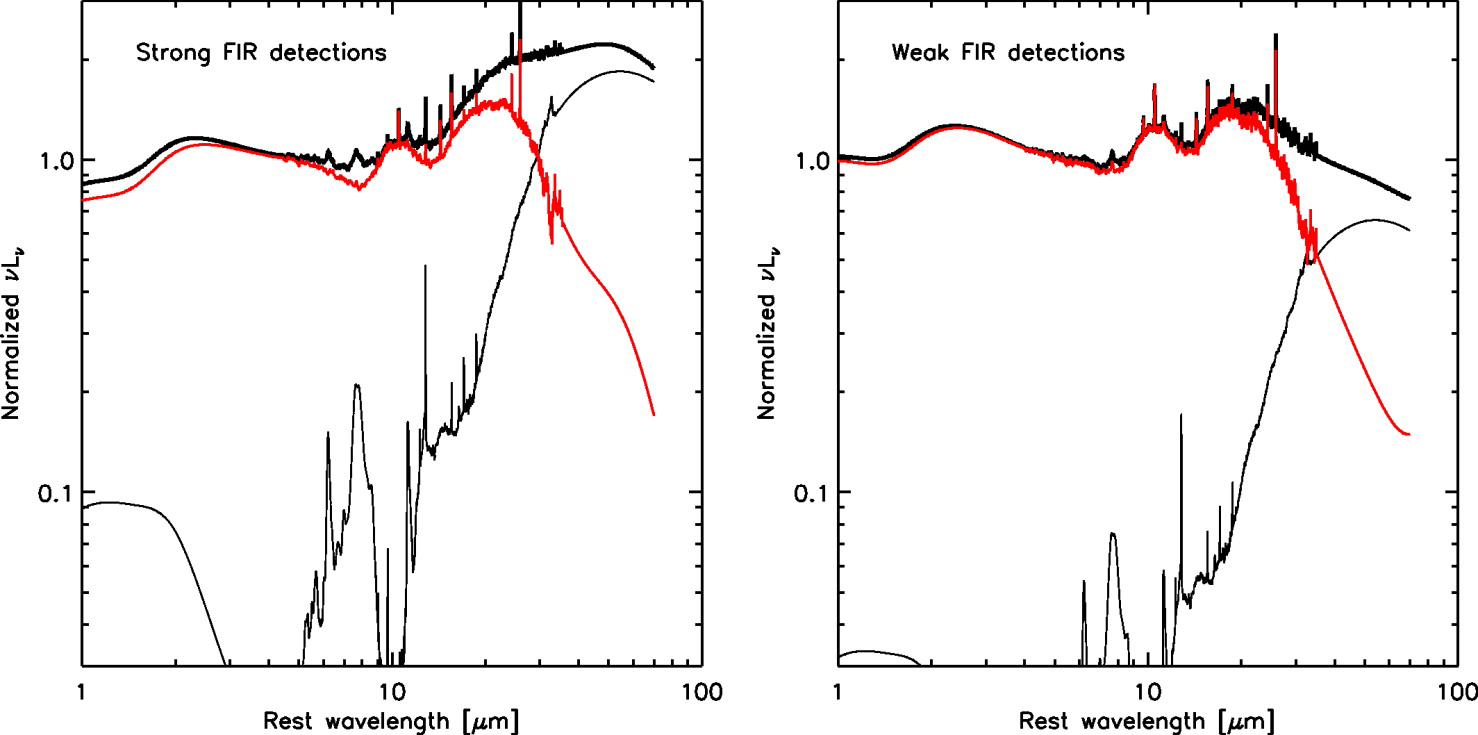}  
%\hspace{5.1cm}
%\end{tabular}
\end{center}
\caption 
{ \label{fig:AGN-IR-Peak}
Normalized mean SEDs for strong FIR quasars (left, top curve) and weak FIR quasars
(right, top curve)\cite{NetzerEtal07} The red SED curves show ``intrinsic" AGN SEDs obtained by the subtraction of the scaled mean starburst (ULIRG) spectrum (shown in black) from the mean SEDs. The red AGN curves indicates how IR emission ($>$1 $\mu$m) carries a substantial portion of the AGN’s energy and that the peak AGN emission is between 20 and 30 $\mu$m. In the PRIMA 25–30 $\mu$m wavelength region, the variability will be dominated by the AGN, regardless of the more constant mid-IR light from the host galaxy, allowing for RM to provide critical insight into the dust distribution around the AGN} 
\end{figure}

Since the IR universally carries a substantial fraction of the total AGN energy output\cite{NetzerEtal07}, Figure \ref{fig:AGN-IR-Peak}, dust absorption plays a fundamental role in the structure and appearance of all AGN. This obscuration also should play a key role in how the energetics of the accretion disk affect its host galaxy which is an expectation that arises from the relationship of the SMBH mass to the bulge mass of the host galaxy\cite{Gebhardt.etal.00,Ferrarese.etal.00}. Finally the dust acts as a calorimeter and measures the luminosity of the accretion disk, both by its overall SED, as well as by showing where the dust sublimation radius is located: the more luminous the accretion disk, the further out the dust sublimation radius, and hence the later the reverberation of the IR wavelengths vs the UV/optical. This luminosity measurement has the potential to be used as a distance indicator independent of the cosmic distance ladder\cite{Yoshii.etal.14}, since the luminosity of the disk is dependent on the temperature profile of the disk and its size, and so is independent of other standard candles such as Cepheids or Supernovae.

Although the dusty torus model has successfully explained many aspects of AGN phenomenology (e.g., polarization\cite{Antonucci&Miller85}), direct observational constraints on its size and structure are difficult to observe because at the distance of even nearby AGN the dust structure is very hard to resolve. But in recent years with the advancements in high angular resolution imaging and interferometry from 8-10m telescopes, especially the advent of the GRAVITY Very Large Telescope Interferometer (VLTI)\cite{GRAVITY2017}, a better picture of the dust distribution has started to emerge where the dusty torus is not the sole dust structure at the center of an AGN but exists in combination with a dusty polar outflow\cite{Hoenig.etal.2012,Hoenig.etal.2013,Lopez-Gonzaga.etal.2014,Tristram.etal.2014,Lopez-Gonzaga.etal.2016,Leftly.etal.2018, Hoenig2019,Leftley.etal.2021}. The origin of this polar outflow is likely a wind that raises dust from the outer surface of the torus into a hollow cone\cite{Kishimoto.Hoenig.2017,Stalevski.etal.2017,Hoenig2019} (Figure \ref{fig:AGN-Schematic}).

A critical aspect of this picture has come about because of the {\it difference} in the temperatures probed by the near-IR ($\sim$1000 - 1500 K) and the mid-IR (A few $\sim$100 K) where the hotter dust close to the dust sublimation temperature shows a disklike structure while the cooler dust shows an extended structure\cite{Hoenig2019,Leftley.etal.2021} In order to put together a definitive picture of how dust is distributed around the accretion disk of an AGN, a combination of both near- and mid- infrared observations will be needed to flesh out the details of the overall dust distribution.

Dust RM has been done exclusively in the near-IR (1-5 $\mu$m) which traces the distance to the inner edge of the dust near the dust sublimation radius\cite{Clavel..etal..1989,Nelson..1996,Koshida.etal.14,Pozo-Nunez.etal.15,Minezaki.etal.19, Landt.etal.2019, Mandal.etal.21}: a distance of a few light weeks for
low-luminosity AGN to a few light months/years for quasars\cite{Koshida.etal.14}. These results have been backed-up by near-IR interferometry\cite{Gravity.Collaboration.2024}. Interferometry from 8–12 $\mu$m has traced dust on much larger scales: tens to hundreds of light years\cite{Burtscher.etal.2013,Lopez-Gonzaga.etal.16,Garci-Bernete.etal.2016}. So, there is a critical gap of knowledge about the distribution of dust between the sublimation radius and the distances probed by the current capabilities of interferometry especially the region where the polar dust arises. Hence by studying the key 25-30 $\mu$m emitting region that is the peak of the emission from dust surrounding the accretion disk (Figure \ref{fig:AGN-IR-Peak}), in combination with near-IR data tracing the dust sublimation radius, the following key questions can be addressed:

\begin{enumerate}
    \item What is the shape of the dust distribution at the center of an AGN: The geometry and density of the Disk and Polar Outflow
    \item How far outward does that dust extend?
    \item How much of the mid-IR emission originates in the dusty torus which is closer to the accretion disk and so should respond to variations quickly, and how much of the mid-IR emission arises in the polar outflow which is further from the accretion disk, and so should respond to variations slowly?
    \item What accretion disk luminosity irradiating the modeled dust distribution would lead to the observed IR spectral energy distribution and dust sublimation radius?
\end{enumerate}

In this paper we present how the proposed PRobe far-Infrared Mission for Astrophysics (PRIMA) will help answer the above questions.

\section{Dust Modeling: Substituting Time Resolution for Spatial Resolution}

\begin{figure}
\begin{center}
%\begin{tabular}{c}
\includegraphics[height=6cm]{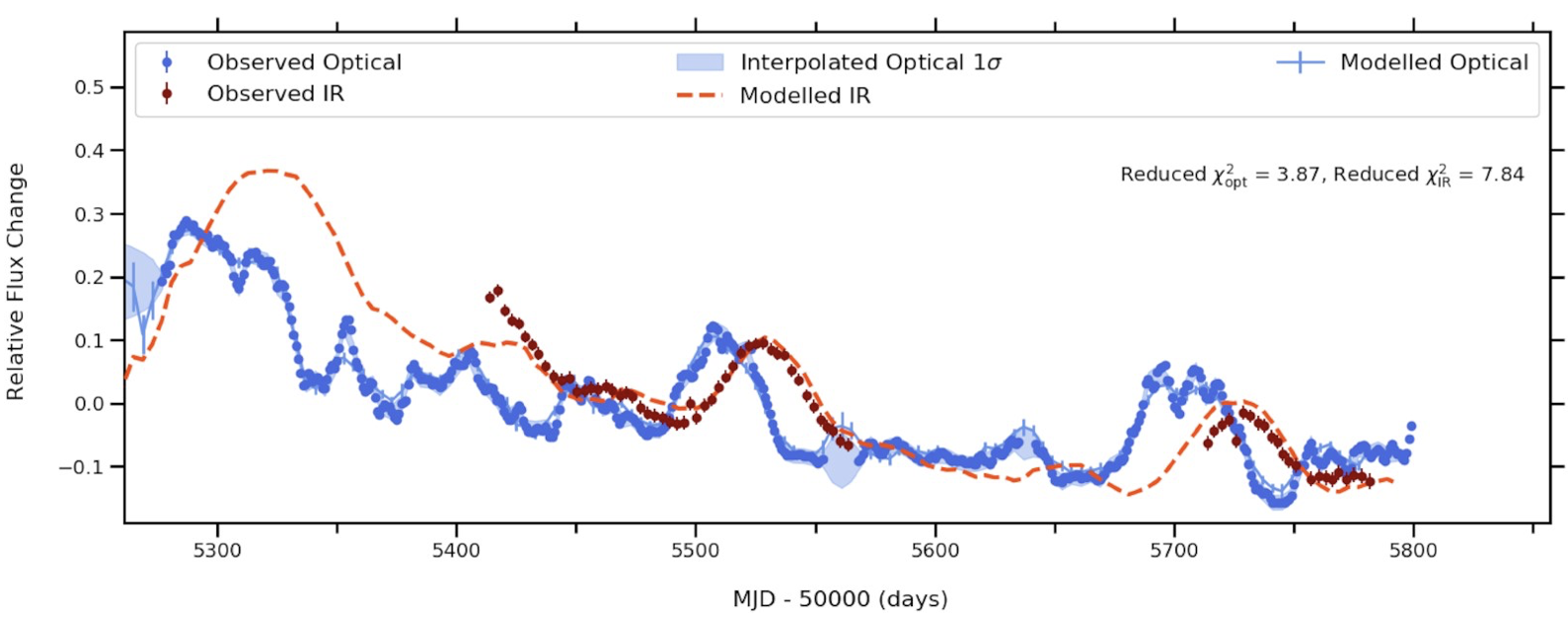}  
%\hspace{5.1cm}
%\end{tabular}
\end{center}
\caption 
{ \label{fig:Zw229-light-curve}
Kepler (dark blue points) and Spitzer (dark red points) observed light curves for the 2010-
2011 observing season for the AGN Zw229-015, plotted with the interpolated optical (light blue) and interpolated
infrared (light red dashes) that corresponded to the highest posterior distribution.}
\end{figure}

Since there are no examples of 25–30 $\mu$m RM, we present here an example of how optical and near-IR RM can be used to understand the dust distribution around the accretion disk of a SMBH at the center of an AGN; thereby illustrating how time resolution can be used in place of spatial resolution. The following two light curves, Figure \ref{fig:Zw229-light-curve}, were used to determine the inner dust distribution of the AGN Zw229-015. We show the optical light curve obtained for the Type I AGN Zw229-015 by the Kepler Space Telescope and the 3.6 $\mu$m light curve provided
by the Spitzer Space Telescope\cite{Guise.etal.22}. As can be seen, the dust reverberated light curve is a time-displaced, smoothed, and widened version of the optical light curve. Each of these differences from the optical light curve to the IR light curve is a function of the dust distribution.

\begin{figure}
\begin{center}
%\begin{tabular}{c}
\includegraphics[height=8cm]{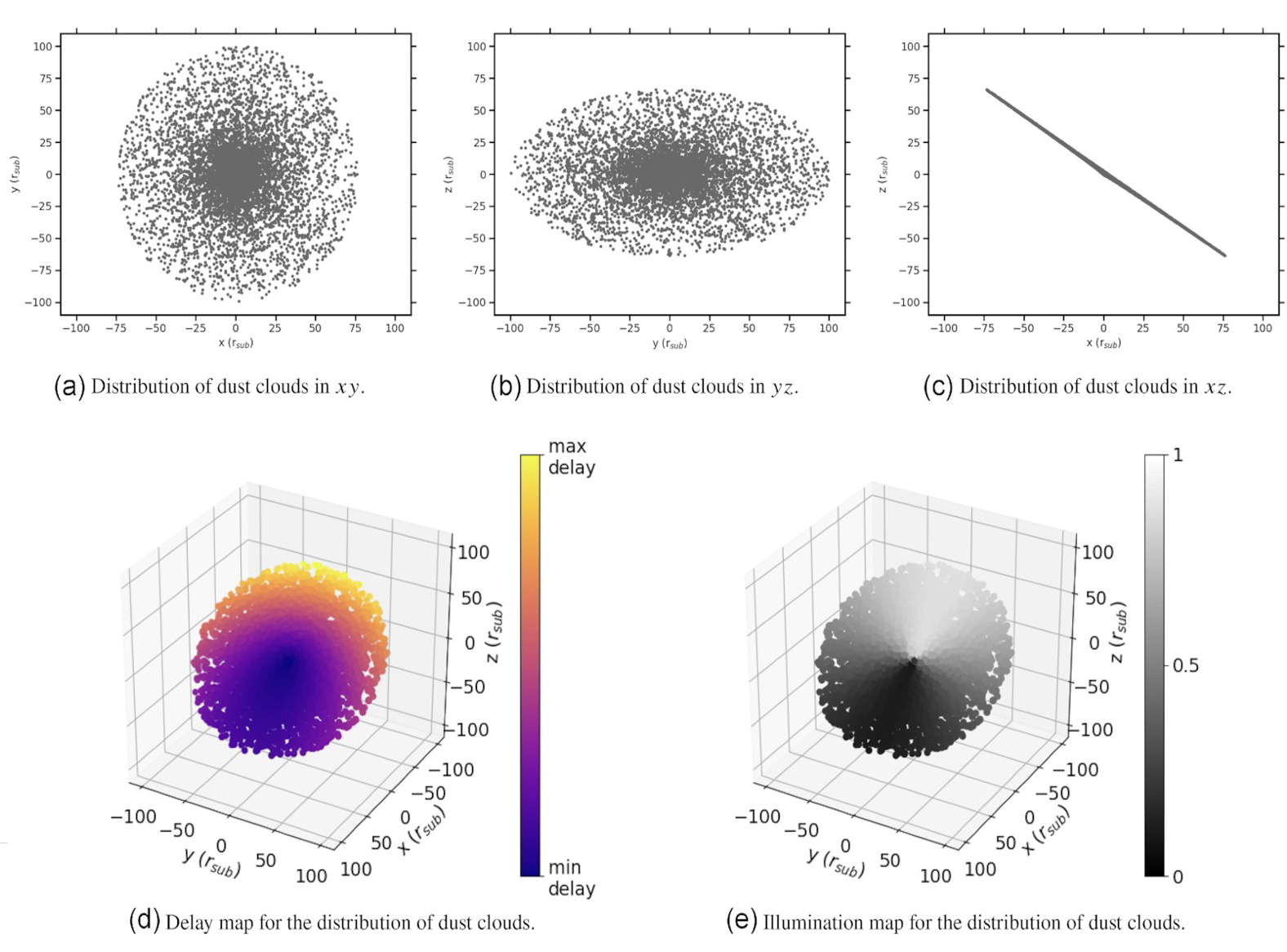}  
%\hspace{5.1cm}
%\end{tabular}
\end{center}
\caption 
{ \label{fig:Zw229-dust-distribtuion}
Distribution of 10,000 dust clouds surrounding the Zw229-015 accretion disk\cite{Guise.etal.22} These distributions reproduce the delayed infrared light curves and are derived from the mean parameters of the MCMC modeling of the delayed IR light curves vs the optical light curves shown in Figure \ref{fig:Zw229-light-curve}. Images {\bf a} , {\bf b} , and {\bf c} show the modeled distribution of clouds from multiple angles. Image {\bf d} shows the modeled delay map of arrival times from the near side and far side of the dust distribution while deriving an inclination of 49$^{+3}_{-13}$ degrees. Finally image {\bf e} shows the derived illumination map of the dust distribution}
\end{figure}

Using MCMC approaches, a model IR light curve at 3.6 $\mu$m is created that best matches a delayed, smoothed, and widened version of the optical light curve\cite{Guise.etal.22}. The displacement between their peaks and valleys gives the first order separation of the 3.6 $\mu$m IR emitting dust from the optical emitting accretion disk, which in this case was found to be 18.3$\pm$4.5 light days at an inclination to our line of sight of 49$^{+3}_{-13}$ degrees.
The light curves are then used to create a dust-cloud distribution model that reproduces the reverberated light curves. In the case of the Zw229-015, a distribution of 10,000 dust clouds corresponding to the mean parameters from the MCMC modeling of the Zw229-015 light curves produced the dust distribution represented in Figure \ref{fig:Zw229-dust-distribtuion}. It must be noted that there are other modeling approaches, such as the TORMAC forward modeling code, that simulates IR reverberation uses by using dust radiation transfer models to simulate IR reverberation at multiple wavelengths\cite{Almeyda.etal.2017,Almeyda.etal.2020}

The above derived dust distribution, particularly the uncertainty in the dust sublimation radius, is sensitive to gaps in the light curve because the modeling is dependent upon inflections in the light curve, and the fewer the inflections, then the harder it is to match the shorter wavelength observations to the longer wavelength observations. In addition to having a cadence that minimizes the gaps, higher sensitivity allows for an increase in precision because smaller inflections (which are more common) are captured giving the matching algorithms more regions between the available light curves to correlate against each other. Thus, the fewer the gaps and the higher the precision, the better and more precise the results. The needed cadence and sensitivity described above are key features of the proposed PRIMA mission.

\section{PRIMA and Reverberation Mapping of AGN}

PRIMA features a cryogenically cooled 1.8 m diameter telescope and is designed to carry two science instruments, a photometer (PRIMAger) and a spectrometer (FIRESS), enabling background limited imaging and spectroscopic studies in the 24 to 235 $\mu$m wavelength range (for more details please see earlier papers in this special issue). PRIMA's capabilities are unmatched for RM. Previously, the Spitzer Space Telescope\cite{Werner.etal.04} could observe at similar wavelengths during its cryogenic mission from 2003 to 2009, but due to its rotating schedule for its 24 $\mu$m camera versus its other instruments, it could not provide the needed regular cadence to do the reverberation observations. The next most recent space infrared mission which had long wavelength infrared capability was the Wide-Field Infrared Survey Explorer (WISE) mission\cite{Wright.etal.10}, which had a 22 $\mu$m channel during its 2009 to 2010 cryogenic mission, but it did not survey the sky long enough to pick up the longe term variations. Currently the James Webb Space Telescope has the needed 25 $\mu$m capability, but due to its large overhead needed to slew to any target, the hundreds of observations needed to provide the cadence to detect reverberation would dwarf the actual on-target observing time, thus making for a radically inefficient use of the telescope, especially when trying to monitor a statistically large sample of AGN ($>$10). PRIMA with its high sensitivity at 25 to 30 $\mu$m, but smaller physical size, will allow for RM programs to repeatedly visit multiple targets without excessive overhead, and get the needed statistics for a large sample of AGN over the five-year lifetime of the mission.

\subsection{Sensitivity}
As noted earlier, good cadence is vital for proper RM and the cadence is a function of the sensitivity of PRIMA. Here we do a preliminary exercise of determining the needed exposure time for both of PRIMA's instruments. {\it Both instruments will need to be used to achieve the needed cadence because they can not be used simultaneously and so will be used in alternating campaigns.} 

\begin{figure}
\begin{center}
%\begin{tabular}{c}
\includegraphics[height=8cm]{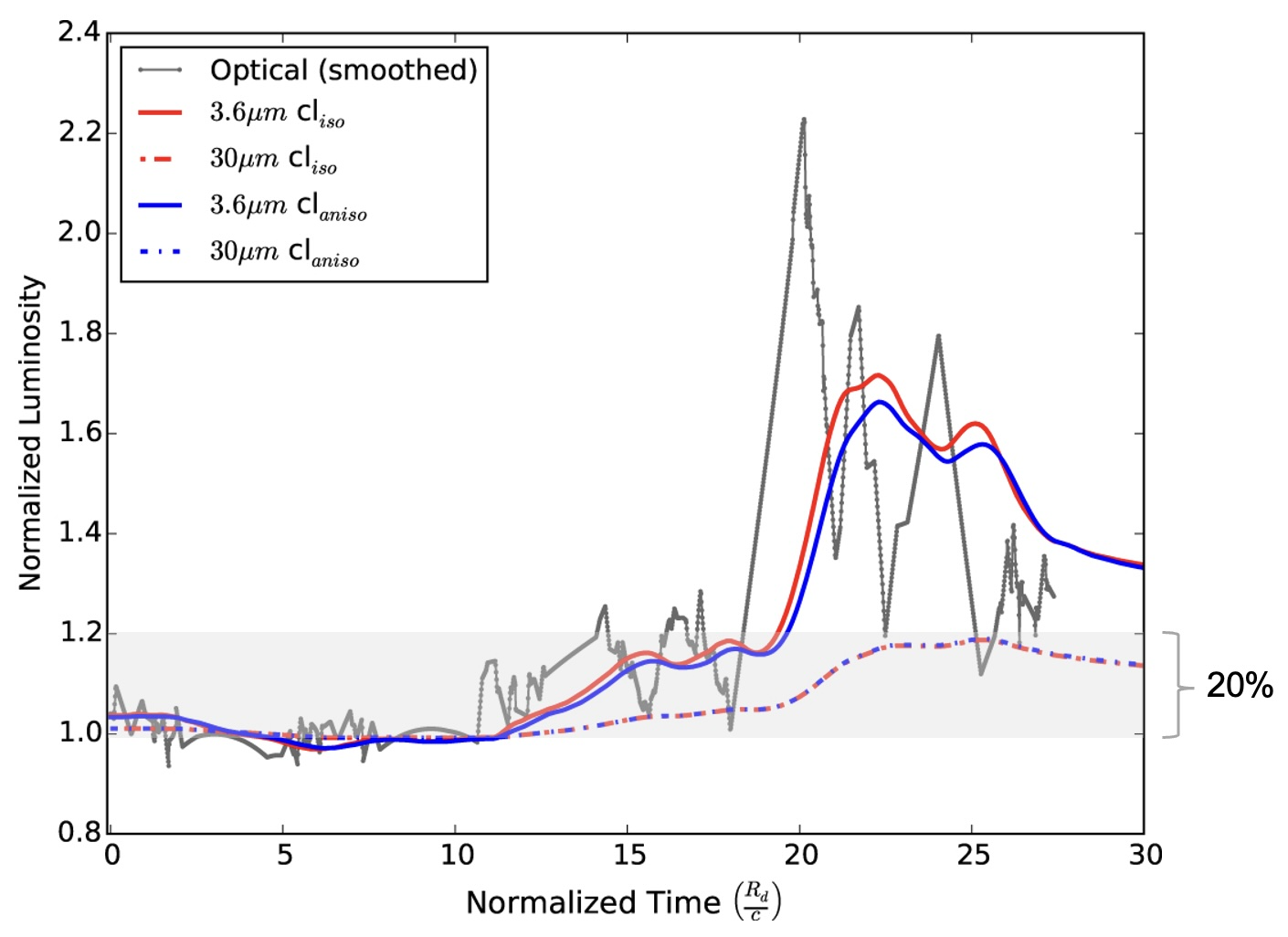}  
%\hspace{5.1cm}
%\end{tabular}
\end{center}
\caption 
{ \label{fig:30micron-RM}
Presenting the observed optical light curve for NGC 6418 with interpolation using a damped random-walk model (gray points and gray solid line) and then simulating the IR response light curves (solid = 3.6 $\mu$m and dashed–dotted = 30 $\mu$m lines). The figure is a slightly modified version from the original publication\cite{Almeyda.etal.2017}. The peak-to-peak variation at 30 $\mu$m is 20\% noted in the gray shaded region on the plot and on the right from which we derive our exposure times. The red line is a simulation of an isotropically illuminated torus where cloud orientations have not been taken into account. The blue line is the same simulation but with the orientation of the clouds taken into account. The torus model is made of 50,000 clouds and is viewed face-on with the ratio of 10 between the inner radius to the outer radius. Finally the power law index of the radial cloud distribution is 0, with the an angular size of 45$^{\circ}$ for the torus\cite{Nenkova.etal.2008}.}
\end{figure} 

PRIMA will not be used to find these AGN, but these are known AGN that have been previously monitored for RM and have known flux densities in the range $\sim$100 to 800 mJy at 25 $\mu$m. But signal to noise needs to be very high to allow for detection of enough inflection points to allow good cross-correlation with shorter wavelengths. To determine what the needed exposure time will be, we used the modeled RM light curve derived from the optical light curve of the AGN NGC 6418 using the TORMAC code. \cite{Almeyda.etal.2017}. The IR light curves reproduced in Figure \ref{fig:30micron-RM} once again show how the UV/optical light curve is delayed, smoothed, and widened by the dust reverberation, more so at 30 $\mu$m than at 3.6 $\mu$m. The modeled 30 $\mu$m light curve shows a peak-to-peak variation of only $\sim$ 20\% and inflections on the order of 1\%. So for our faintest source of 100 mJy, those 1\% fluctuations would translate into 1 mJy changes and so we need to be able to resolve changes smaller than 1 mJy and so need to detect at least 0.3mJy changes at high SNR. Note that the actual flux at 25 $\mu$m for NGC 6418 is 12 mJy, and so, by setting our faint source limit at 100 mJy, we will be monitoring sources that are much brighter than NGC 6418 resulting in much better signal to noise.

So according to the current PRIMA Exposure Time Calculator\footnote{https://prima.ipac.caltech.edu/page/etc-calc} for the PRIMAger, to get 5$\sigma$ SNR at 0.3 mJy with a small mosaic (10 sq. arcmin.) will take a $\sim$30 minutes of integration. For the same source precision but observed with FIRESS, the exposure times will be shorter since FIRESS does not require any mosaicing. Thus taking a FIRESS spectrum in its low resolution mode with a resolving power of R=130, and then combining the spectral bins by a factor of 13 to give the same resolving power as PRIMAger, requires an exposure time of $\sim$10 minutes. In both cases there are still uncertainties about the exact exposure time, but it is safe to say that even for this faintest source case, we expect to get the required SNR in approximately half an hour or less, thus giving both the sensitivity and the efficiency to achieve the needed cadence.

\begin{figure}
\begin{center}
%\begin{tabular}{c}
\includegraphics[height=8cm]{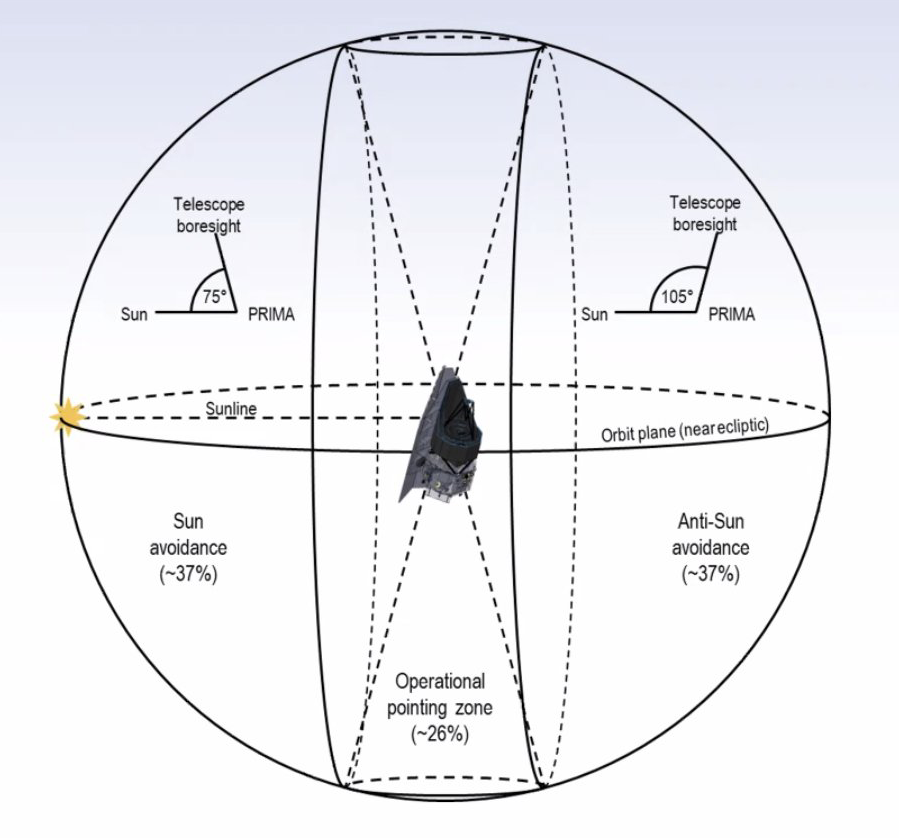}  
%\hspace{5.1cm}
%\end{tabular}
\end{center}
\caption 
{ \label{fig:PRIMA-Field-of-Regard}
PRIMA field of regard $\pm15$ degrees from plane normal to the Sun-Earth line. As the orbit
progresses during the year, the field of regard covers the entire sky with the North and South Ecliptic poles getting continuous coverage.}
\end{figure} 

\subsection{Target Visibility}
Although the exposure time estimation shows that the cadence needed does not tax the observatory's capabilities, there is a final necessity and that is the targets need to be visible to the observatory on a near continuous basis. This need restricts sources to be placed near the North and South Ecliptic poles of PRIMA's Sun-Earth L2 orbit, Figure \ref{fig:PRIMA-Field-of-Regard}, but due to PRIMA's 30 degree field of regard on the sky around the poles, a large swath of sky and so a large number of sources are available to monitor, from space as well as on the ground.

As the Spitzer Space Telescope had a similar restriction, our experience showed that for AGN like Mrk 817 and Zw229-015, we could get over 220 days of continuous observations. These are ideal candidates because they are low luminosity AGN and so their 25-30 $\mu$m emitting region will be closer to the accretion disk and so their reverberation timescales will be shorter (on the order of weeks), and so can be easily captured in those 220 days.

For more luminous AGN like quasars, the reverberation timescale would be much longer, and so would need a correspondingly longer continuous timespan. Fortunately, within the continuous viewing zone (CVZ) near the North and South Ecliptic Poles, there are multiple quasars which would lend themselves to long term monitoring lasting for a few years to detect reverberation. As a first order estimate, using the AGN catalog compiled from the Wide-Field Infrared Survey Explorer (WISE)\cite{Assef.etal.2018}, there are 10 AGN brighter than 100 mJy in the northern CVZ. A similar number should exist in the southern CVZ but are not included in the WISE AGN catalog because the region overlaps with the Large Magellanic Cloud (LMC) which is excluded from the catalog. We will begin preliminary efforts to monitor these sources in the optical and near-IR to identify those AGN that show the greatest variability, especially ones that show wide optical flux variations like those in figure \ref{fig:30micron-RM}, and so would be best to monitor once PRIMA launches.

These targets would be identified before PRIMA's launch, and then they would be placed within the observing plan of PRIMA such that each source would be revisited once every 4-5 days as that would cover the expected delay and smoothing of the optical light curve. Since these observations don’t have to be taken at an exact moment in time, a buffer of $\pm1$ days will be added to the timing. This provides the PRIMA schedulers enough slack so that these repeated observations can be optimally placed in the schedule to take the least amount of slew-overhead as well as not interrupting other extended observations.

This continuous and near continuous monitoring is made possible by using both the spectrometer and the photometer for the variability monitoring, both of which have wavelength coverage from 25-30 $\mu$m\footnote{The two instruments will be cross calibrated at their overlapping wavelengths}, and so the cadence will not be compromised when switching campaigns between the instruments. The expected lag between the optical and the 25-30 $\mu$m emission region can be long (months to years), but the nominal five-year duration of the PRIMA mission will allow for reverberation to be detected.

\subsection{Synergies with Other Observatories}

Since RM observations for dust reverberation are dependent on monitoring short wavelength emission
from the accretion disk, optical observations are necessary. Fortunately, with the advent of large-scale
optical monitoring surveys from the ground (e.g., Zwicky Transient Facility, Vera C. Rubin Observatory) that predate the PRIMA launch (expected to be around 2032), there will not
be any need to plan for separate optical observations.
But it is always beneficial to have additional information, especially in the near-IR as there will not be the same level of automated observations as in the optical. To address this, the AGN RM community has a
consortium that gets multiwavelength photometric and spectroscopic monitoring for any AGN that is
being monitored for reverberation, based on the availability of the facilities that each member has access
to. Members also apply to general observer facilities on the ground and in space to supplement the RM
data. AGN that are being monitored from space get a higher priority from the team members. This
process has been employed successfully for decades starting with observations by the International
Ultraviolet Explorer of six AGN from 1988 to 1996\cite{Peterson92} and continuing till today with observations by the AGN Space Telescope and Optical Reverberation Mapping (AGN STORM) team of NGC 5548 and Mrk 817
led by the Hubble Space Telescope \cite{derosa.etal.15,kara.etal.21} and supplemented by
multiple ground and space based telescopes like LCO, Swift, XMM, IRTF, and others. We expect that for a future PRIMA time allocation committee to accept a RM proposal, a proposing team would be able to show that they had already assembled a team capable of getting the additional supporting data from the ground and from space similar to those noted above but especially near infrared resources from Palomar, Keck, Gemini, and the VLT to trace the hottest part of the dust emission.

\section{Summary and Conclusions}

Understanding the dust distribution around the accretion disks of AGN is a critical part of understanding both the energetics of AGN and the basic structure of accretion onto the supermassive black hole at the center of an AGN. The development of RM over the past $\sim$40 years has already given us more information about many aspects of AGN from the BLR cloud distribution to the mass of the central black hole to the dust sublimation radius. The promise of the PRIMA mission is that we can significantly expand our understanding of the {\it overall} dust around AGN and thereby increase our understanding of AGN. This is because the dust distribution is vitally connected to the structure of the AGN from the role it plays in the Type I vs Type II classification to how it shields and directs the AGN emission into the host galaxy. All this in addition to the fact that the dust luminosity is directly related to the accretion disk luminosity which may eventually be used as a standard candle.

These key contributions by PRIMA are not only facilitated by its high sensitivity at the peak of the dust emission of AGN in the 25-30 $\mu$m range, but also by its ability to continuously observe at those wavelengths with either its photometer and spectrometer, thus providing the vital cadence necessary for getting properly matched multiwavelength light curves.

Another fortuitous aspect for RM with PRIMA is that a problem that has plagued RM science from the very beginning has been solved and that is the availability of high cadence observations at other wavelengths. Even before the potential launch of PRIMA, shorter optical wavelength monitoring of large numbers of AGN will be provided by large scale time-domain facilities like the Zwicky Transient Facility and the Vera C. Rubin Observatory. Additionally the AGN RM community routinely supports space based RM observation campaigns by providing additional observations. 

All of the above advantages create the landscape for PRIMA to make vital contributions to AGN science.

% \disclosures 
\subsection*{Disclosures}
The authors are both affiliated with the Jet Propulsion Laboratory/California Institute of Technology which is one of the centers that is connected with the PRIMA step 2 proposal, though neither is funded by NASA Step 2 funding.

\subsection* {Code, Data, and Materials Availability} 
Data sharing is not applicable to this article, as no new data were created or analyzed.

\subsection* {Acknowledgments}
The research was carried out at the Jet Propulsion Laboratory, California Institute of Technology, under a contract with the National Aeronautics and Space Administration (80NM0018D0004).. This work is based in part on observations made with the Spitzer Space Telescope, which was operated by the Jet Propulsion Laboratory, California Institute of Technology under a contract with NASA.

%%%%% References %%%%%

\bibliography{report}   % bibliography data in report.bib

\begin{thebibliography}{10}

\bibitem{Blandford..1982}
R.~D. {Blandford} and C.~F. {McKee}, ``{Reverberation mapping of the emission line regions of Seyfert galaxies and quasars.},'' {\em ApJ} {\bf 255}, 419--439  (1982).

\bibitem{Peterson..1993}
B.~M. {Peterson}, ``{Reverberation Mapping of Active Galactic Nuclei},'' {\em PASP} {\bf 105}, 247  (1993).

\bibitem{Kriss.etal.2000}
G.~A. {Kriss}, B.~M. {Peterson}, D.~M. {Crenshaw}, {\em et~al.}, ``{A High Signal-to-Noise Ultraviolet Spectrum of NGC 7469: New Support for Reprocessing of Continuum Radiation},'' {\em \apj} {\bf 535}, 58--72  (2000).

\bibitem{Sergeev.etal.2005}
S.~G. {Sergeev}, V.~T. {Doroshenko}, Y.~V. {Golubinskiy}, {\em et~al.}, ``{Lag-Luminosity Relationship for Interband Lags between Variations in B, V, R, and I Bands in Active Galactic Nuclei},'' {\em \apj} {\bf 622}, 129--135  (2005).

\bibitem{Shappee.etal.2014}
B.~J. {Shappee}, J.~L. {Prieto}, D.~{Grupe}, {\em et~al.}, ``{The Man behind the Curtain: X-Rays Drive the UV through NIR Variability in the 2013 Active Galactic Nucleus Outburst in NGC 2617},'' {\em \apj} {\bf 788}, 48  (2014).

\bibitem{Clavel..etal..1989}
J.~{Clavel}, W.~{Wamsteker}, and I.~S. {Glass}, ``{Hot Dust on the Outskirts of the Broad-Line Region in Fairall 9},'' {\em ApJ} {\bf 337}, 236  (1989).

\bibitem{Nelson..1996}
B.~O. {Nelson}, ``{A Correlated Optical-Infrared Outburst of Markarian 744: The Strongest Evidence Yet for Thermal Dust Reverberation},'' {\em ApJL} {\bf 465}, L87  (1996).

\bibitem{Antonucci93}
R.~{Antonucci}, ``{Unified models for active galactic nuclei and quasars.},'' {\em ARA\&A} {\bf 31}, 473--521  (1993).

\bibitem{CackettBentzKara2021}
E.~M. {Cackett}, M.~C. {Bentz}, and E.~{Kara}, ``{Reverberation mapping of active galactic nuclei: from X-ray corona to dusty torus},'' {\em iScience} {\bf 24}, 102557  (2021).

\bibitem{NetzerEtal07}
H.~{Netzer}, D.~{Lutz}, M.~{Schweitzer}, {\em et~al.}, ``{Spitzer Quasar and ULIRG Evolution Study (QUEST). II. The Spectral Energy Distributions of Palomar-Green Quasars},'' {\em ApJ} {\bf 666}, 806--816  (2007).

\bibitem{Gebhardt.etal.00}
K.~{Gebhardt}, R.~{Bender}, G.~{Bower}, {\em et~al.}, ``{A Relationship between Nuclear Black Hole Mass and Galaxy Velocity Dispersion},'' {\em ApJL} {\bf 539}, L13--L16  (2000).

\bibitem{Ferrarese.etal.00}
L.~{Ferrarese} and D.~{Merritt}, ``{A Fundamental Relation between Supermassive Black Holes and Their Host Galaxies},'' {\em ApJL} {\bf 539}, L9--L12  (2000).

\bibitem{Yoshii.etal.14}
Y.~{Yoshii}, Y.~{Kobayashi}, T.~{Minezaki}, {\em et~al.}, ``{A New Method for Measuring Extragalactic Distances},'' {\em ApJL} {\bf 784}, L11  (2014).

\bibitem{Antonucci&Miller85}
R.~R.~J. {Antonucci} and J.~S. {Miller}, ``{Spectropolarimetry and the nature of NGC 1068.},'' {\em ApJ} {\bf 297}, 621--632  (1985).

\bibitem{GRAVITY2017}
{GRAVITY Collaboration}, R.~{Abuter}, M.~{Accardo}, {\em et~al.}, ``{First light for GRAVITY: Phase referencing optical interferometry for the Very Large Telescope Interferometer},'' {\em \aap} {\bf 602}, A94  (2017).

\bibitem{Hoenig.etal.2012}
S.~F. {H{\"o}nig}, M.~{Kishimoto}, R.~{Antonucci}, {\em et~al.}, ``{Parsec-scale Dust Emission from the Polar Region in the Type 2 Nucleus of NGC 424},'' {\em \apj} {\bf 755}, 149  (2012).

\bibitem{Hoenig.etal.2013}
S.~F. {H{\"o}nig}, M.~{Kishimoto}, K.~R.~W. {Tristram}, {\em et~al.}, ``{Dust in the Polar Region as a Major Contributor to the Infrared Emission of Active Galactic Nuclei},'' {\em \apj} {\bf 771}, 87  (2013).

\bibitem{Lopez-Gonzaga.etal.2014}
N.~{L{\'o}pez-Gonzaga}, W.~{Jaffe}, L.~{Burtscher}, {\em et~al.}, ``{Revealing the large nuclear dust structures in NGC 1068 with MIDI/VLTI},'' {\em \aap} {\bf 565}, A71  (2014).

\bibitem{Tristram.etal.2014}
K.~R.~W. {Tristram}, L.~{Burtscher}, W.~{Jaffe}, {\em et~al.}, ``{The dusty torus in the Circinus galaxy: a dense disk and the torus funnel},'' {\em \aap} {\bf 563}, A82  (2014).

\bibitem{Lopez-Gonzaga.etal.2016}
N.~{L{\'o}pez-Gonzaga}, L.~{Burtscher}, K.~R.~W. {Tristram}, {\em et~al.}, ``{Mid-infrared interferometry of 23 AGN tori: On the significance of polar-elongated emission},'' {\em \aap} {\bf 591}, A47  (2016).

\bibitem{Leftly.etal.2018}
J.~H. {Leftley}, K.~R.~W. {Tristram}, S.~F. {H{\"o}nig}, {\em et~al.}, ``{New Evidence for the Dusty Wind Model: Polar Dust and a Hot Core in the Type-1 Seyfert ESO 323-G77},'' {\em \apj} {\bf 862}, 17  (2018).

\bibitem{Hoenig2019}
S.~F. {H{\"o}nig}, ``{Redefining the Torus: A Unifying View of AGNs in the Infrared and Submillimeter},'' {\em \apj} {\bf 884}, 171  (2019).

\bibitem{Leftley.etal.2021}
J.~H. {Leftley}, K.~R.~W. {Tristram}, S.~F. {H{\"o}nig}, {\em et~al.}, ``{Resolving the Hot Dust Disk of ESO323-G77},'' {\em \apj} {\bf 912}, 96  (2021).

\bibitem{Kishimoto.Hoenig.2017}
S.~F. {H{\"o}nig} and M.~{Kishimoto}, ``{Dusty Winds in Active Galactic Nuclei: Reconciling Observations with Models},'' {\em \apjl} {\bf 838}, L20  (2017).

\bibitem{Stalevski.etal.2017}
M.~{Stalevski}, D.~{Asmus}, and K.~R.~W. {Tristram}, ``{Dissecting the active galactic nucleus in Circinus - I. Peculiar mid-IR morphology explained by a dusty hollow cone},'' {\em \mnras} {\bf 472}, 3854--3870  (2017).

\bibitem{Koshida.etal.14}
S.~{Koshida}, T.~{Minezaki}, Y.~{Yoshii}, {\em et~al.}, ``{Reverberation Measurements of the Inner Radius of the Dust Torus in 17 Seyfert Galaxies},'' {\em ApJ} {\bf 788}, 159  (2014).

\bibitem{Pozo-Nunez.etal.15}
F.~{Pozo Nu{\~n}ez}, M.~{Ramolla}, C.~{Westhues}, {\em et~al.}, ``{The broad-line region and dust torus size of the Seyfert 1 galaxy PGC 50427},'' {\em A\&A} {\bf 576}, A73  (2015).

\bibitem{Minezaki.etal.19}
T.~{Minezaki}, Y.~{Yoshii}, Y.~{Kobayashi}, {\em et~al.}, ``{Reverberation Measurements of the Inner Radii of the Dust Tori in Quasars},'' {\em ApJ} {\bf 886}, 150  (2019).

\bibitem{Landt.etal.2019}
H.~{Landt}, M.~J. {Ward}, D.~{Kynoch}, {\em et~al.}, ``{The first spectroscopic dust reverberation programme on active galactic nuclei: the torus in NGC 5548},'' {\em \mnras} {\bf 489}, 1572--1589  (2019).

\bibitem{Mandal.etal.21}
A.~K. {Mandal}, S.~{Rakshit}, C.~S. {Stalin}, {\em et~al.}, ``{Dust reverberation mapping of Z229-15},'' {\em MNRAS} {\bf 501}, 3905--3915  (2021).

\bibitem{Gravity.Collaboration.2024}
{Gravity Collaboration}, A.~{Amorim}, G.~{Bourdarot}, {\em et~al.}, ``{VLTI/GRAVITY interferometric measurements of the innermost dust structure sizes around active galactic nuclei},'' {\em \aap} {\bf 690}, A76  (2024).

\bibitem{Burtscher.etal.2013}
L.~{Burtscher} and K.~R.~W. {Tristram}, ``{The Diversity of Dusty AGN Tori: Results from the VLTI/MIDI AGN Large Programme},'' {\em The Messenger} {\bf 154}, 62--65  (2013).

\bibitem{Lopez-Gonzaga.etal.16}
N.~{L{\'o}pez-Gonzaga}, L.~{Burtscher}, K.~R.~W. {Tristram}, {\em et~al.}, ``{Mid-infrared interferometry of 23 AGN tori: On the significance of polar-elongated emission},'' {\em A\&A} {\bf 591}, A47  (2016).

\bibitem{Garci-Bernete.etal.2016}
I.~{Garc{\'\i}a-Bernete}, C.~{Ramos Almeida}, J.~A. {Acosta-Pulido}, {\em et~al.}, ``{The nuclear and extended mid-infrared emission of Seyfert galaxies},'' {\em \mnras} {\bf 463}, 3531--3555  (2016).

\bibitem{Guise.etal.22}
E.~{Guise}, S.~F. {H{\"o}nig}, V.~{Gorjian}, {\em et~al.}, ``{Dust reverberation mapping and light-curve modelling of Zw229-015},'' {\em MNRAS} {\bf 516}, 4898--4915  (2022).

\bibitem{Almeyda.etal.2017}
T.~{Almeyda}, A.~{Robinson}, M.~{Richmond}, {\em et~al.}, ``{Modeling the Infrared Reverberation Response of the Circumnuclear Dusty Torus in AGNs: The Effects of Cloud Orientation and Anisotropic Illumination},'' {\em \apj} {\bf 843}, 3  (2017).

\bibitem{Almeyda.etal.2020}
T.~{Almeyda}, A.~{Robinson}, M.~{Richmond}, {\em et~al.}, ``{Modeling the Infrared Reverberation Response of the Circumnuclear Dusty Torus in AGNs: An Investigation of Torus Response Functions},'' {\em \apj} {\bf 891}, 26  (2020).

\bibitem{Werner.etal.04}
M.~W. {Werner}, T.~L. {Roellig}, F.~J. {Low}, {\em et~al.}, ``{The Spitzer Space Telescope Mission},'' {\em ApJS} {\bf 154}, 1--9  (2004).

\bibitem{Wright.etal.10}
E.~L. {Wright}, P.~R.~M. {Eisenhardt}, A.~K. {Mainzer}, {\em et~al.}, ``{The Wide-field Infrared Survey Explorer (WISE): Mission Description and Initial On-orbit Performance},'' {\em AJ} {\bf 140}, 1868--1881  (2010).

\bibitem{Nenkova.etal.2008}
M.~{Nenkova}, M.~M. {Sirocky}, R.~{Nikutta}, {\em et~al.}, ``{AGN Dusty Tori. II. Observational Implications of Clumpiness},'' {\em \apj} {\bf 685}, 160--180  (2008).

\bibitem{Assef.etal.2018}
R.~J. {Assef}, D.~{Stern}, G.~{Noirot}, {\em et~al.}, ``{The WISE AGN Catalog},'' {\em \apjs} {\bf 234}, 23  (2018).

\bibitem{Peterson92}
B.~M. {Peterson}, ``{Light Echoes in Active Galactic Nuclei},'' in {\em American Astronomical Society Meeting Abstracts \#180},  {\em American Astronomical Society Meeting Abstracts} {\bf 180}, 20.02  (1992).

\bibitem{derosa.etal.15}
G.~{De Rosa}, B.~M. {Peterson}, J.~{Ely}, {\em et~al.}, ``{Space Telescope and Optical Reverberation Mapping Project.I. Ultraviolet Observations of the Seyfert 1 Galaxy NGC 5548 with the Cosmic Origins Spectrograph on Hubble Space Telescope},'' {\em ApJ} {\bf 806}, 128  (2015).

\bibitem{kara.etal.21}
E.~{Kara}, M.~{Mehdipour}, G.~A. {Kriss}, {\em et~al.}, ``{AGN STORM 2. I. First results: A Change in the Weather of Mrk 817},'' {\em ApJ} {\bf 922}, 151  (2021).

\end{thebibliography}
\bibliographystyle{spiejour}   % makes bibtex use spiejour.bst

%%%%% Biographies of authors %%%%%

\vspace{2ex}\noindent\textbf{Dr. Varoujan Gorjian} is a Research Scientist at NASA’s Jet Propulsion Laboratory, California Institute of Technology. He was involved with the Spitzer Space Telescope for over 20 years as both a member of the Spitzer Project Science office at JPL as well as a scientific user of the telescope, in particular to do reverberation mapping of active galactic nuclei (AGN). His main astronomical interests continue to be variability in AGN, as well characterization of exoplanets and their host stars, and detection of the cosmic infrared background.

\vspace{2ex}\noindent\textbf{Dr. Michael Werner} is the Project Scientist of the Spitzer Space Telescope at the Jet Propulsion
Laboratory, California Institute of Technology. He has also served as the Chief Scientist for Astronomy and Physics at JPL. He has authored and coauthored dozens of papers presenting scientific results from Spitzer and is the author, with coauthor Peter Eisenhardt, of the book ``More Things in the
Heavens: How Infrared Astronomy is Expanding our View of the Universe,” (Princeton University Press, 2019), which discusses Spitzer science in depth.

%\vspace{1ex}
%\noindent Biographies and photographs of the other authors are not available.

\listoffigures
%\listoftables

\end{spacing}

\end{document}